\documentclass[a4paper]{jpconf}
\usepackage{graphicx}
\begin{document}
\title{Thermal and non-thermal charmed meson production in heavy ions collisions at the LHC}

\author{I P Lokhtin$^{1,*}$, A V Belyaev$^1$, G Kh Eyyubova$^{1,2}$, G Ponimatkin$^3$ 
and E Yu Pronina$^{1}$ }

\address{$^1$ Skobeltsyn Institute of Nuclear Physics, Lomonosov Moscow
	State University, Moscow, Russia}

\address{$^2$ Faculty of Nuclear Sciences and Physical Engineering, Czech Technical University in 
Prague, Prague, Czech Republic}

\address{$^3$ Ostrov Industrial High School, Ostrov, Karlovy Vary District, Czech Republic}

\ead{Igor.Lokhtin@cern.ch}

\begin{abstract}
The phenomenological analysis of the LHC data on $p_{\rm T}$-spectrum and elliptic flow of 
$J/\psi$ and D mesons in PbPb collisions at $\sqrt{s_{\rm NN}}=2.76$ TeV is presented. The 
charmed meson production pattern in PbPb collisions may be reproduced by two-component model 
HYDJET++  including thermal and non-thermal components. The significant part of 
D-mesons is found to be in a kinetic equilibrium with the created medium, while $J/\psi$-mesons 
are not.
\end{abstract}

\section{Introduction}

Studying the charmed hadron production is a particularly useful tool to probe properties of hot 
and dense matter created in relativistic heavy ion collisions. 
The modern pattern of multi-particle production in (most central) heavy ion collisions at RHIC and 
LHC agrees with the formation of hot strongly-interacting matter with hydrodynamical properties (``quark-gluon fluid''), which absorbs energetic quarks and gluons due to their multiple scattering and medium-induced energy loss (see, e.g.,~\cite{qm14}). Within such paradigm, a number of questions 
on heavy flavor production arises. Are heavy quarks thermalized in quark-gluon plasma? What
is the mass dependence of in-medium quark energy loss? Are charmed hadrons in a
kinetic equilibrium with the medium? How the specific pattern of quarkonium suppression is related to the interplay between thermal and non-thermal mechanisms of $J/\psi$-meson production? In this paper, some results of the phenomenological analysis of LHC data on momentum spectra and elliptic flow of charmed
hadrons (D, $J/\psi$) in PbPb collisions at $\sqrt{s_{\rm NN}}=2.76$ TeV within 
two-component model HYDJET++ are presented. The comparison with RHIC results is also discussed. 

\section{HYDJET++ model}

HYDJET++ is the model of relativistic heavy ion collisions, which includes two independent 
components: the soft hydro-type state (``thermal'' component) and the hard state resulting from 
the medium-modified multi-parton fragmentation (``non-thermal'' component)~\cite{Lokhtin:2008xi}. 

The soft component represents the hadronic state generated on the chemical and thermal freeze-out hypersurfaces obtained from the pa\-ra\-met\-ri\-za\-ti\-on of relativistic hydrodynamics with 
preset freeze-out conditions (the adapted event generator FAST 
MC~\cite{Amelin:2006qe,Amelin:2007ic}). Hadron multiplicities are calculated using the effective thermal volume approximation and Poisson multiplicity distribution around its mean value, 
which is supposed to be proportional to a number of participating nucleons for a given impact
parameter of an AA collision. To simulate the elliptic flow effect, the hydro-inspired 
pa\-ra\-met\-ri\-za\-ti\-on is implemented for the momentum and spatial anisotropy of a thermal 
hadron emission source.

The approach used for the hard component is based on the PYQUEN partonic energy loss 
model~\cite{Lokhtin:2005px}. The simulation of single hard NN sub-collision by PYQUEN is 
constructed as a modification of the jet event obtained with PYTHIA$\_$6.4  
generator~\cite{Sjostrand:2006za}. It takes into account medium-induced rescattering, collisional 
and radiative energy loss of hard partons in the expanding quark-gluon fluid, realistic nuclear
geometry and nuclear shadowing effect. The mean number of jets produced in an AA event is calculated as a product of the number of binary NN sub-collisions at a given impact parameter per the integral
cross section of the hard process with the minimum transverse momentum transfer $p_{\rm T}^{\rm min}$. The partons produced in hard processes with the momentum transfer lower
than $p_{\rm T}^{\rm min}$, are considered as being ``thermalized''. So, their hadronization
products are included ``automatically'' in the soft component of the event. 

The input parameters of the model for soft and hard components have been tuned from fitting to heavy ion data on various observables for inclusive hadrons at RHIC~\cite{Lokhtin:2008xi} and LHC~\cite{Lokhtin:2012re}. 

Charmed meson production in HYDJET++ includes soft and hard components as well. Thermal production 
of $D$, $J/\psi$ and $\Lambda_c$ hadrons is treated within the statistical hadronization 
approach~\cite{Andronic:2003zv,Andronic:2006ky}. Momentum spectra of charm hadrons are computed
according to the thermal distribution, and the multiplicities $N_c$ ($C = D, J/\psi, \Lambda_c$) are
calculated through the corresponding thermal numbers $N_c^{\rm th}$ as $N_c=\gamma_c^{\rm n_c}  
N_c^{\rm th}$, where $\gamma_c$ is the charm enhancement factor (or charm fugacity), and $n_c$ is the number of charm quarks in a hadron $C$. The fugacity $\gamma_c$ can be treated as a free parameter of the model, or calculated through the number of charm quark pairs obtained from PYTHIA and multiplied
by the number of NN sub-collisions. Non-thermal charmed hadrons are generated by PYQUEN taking into account in-medium energy loss of heavy (b, c) quarks 
within ``dead-cone'' generalization~\cite{Dokshitzer:2001zm} of BDMPS 
model~\cite{Baier:1996kr,Baier:1999ds} for radiative loss and high-momentum transfer 
limit~\cite{Bjorken:1982tu,Braaten:1991jj,Lokhtin:2000wm} for collisional loss. 

\section{$J/\psi$-meson production in PbPb collisions at the LHC}

\begin{figure}[h]
\begin{minipage}{18pc}
\includegraphics[width=18pc]{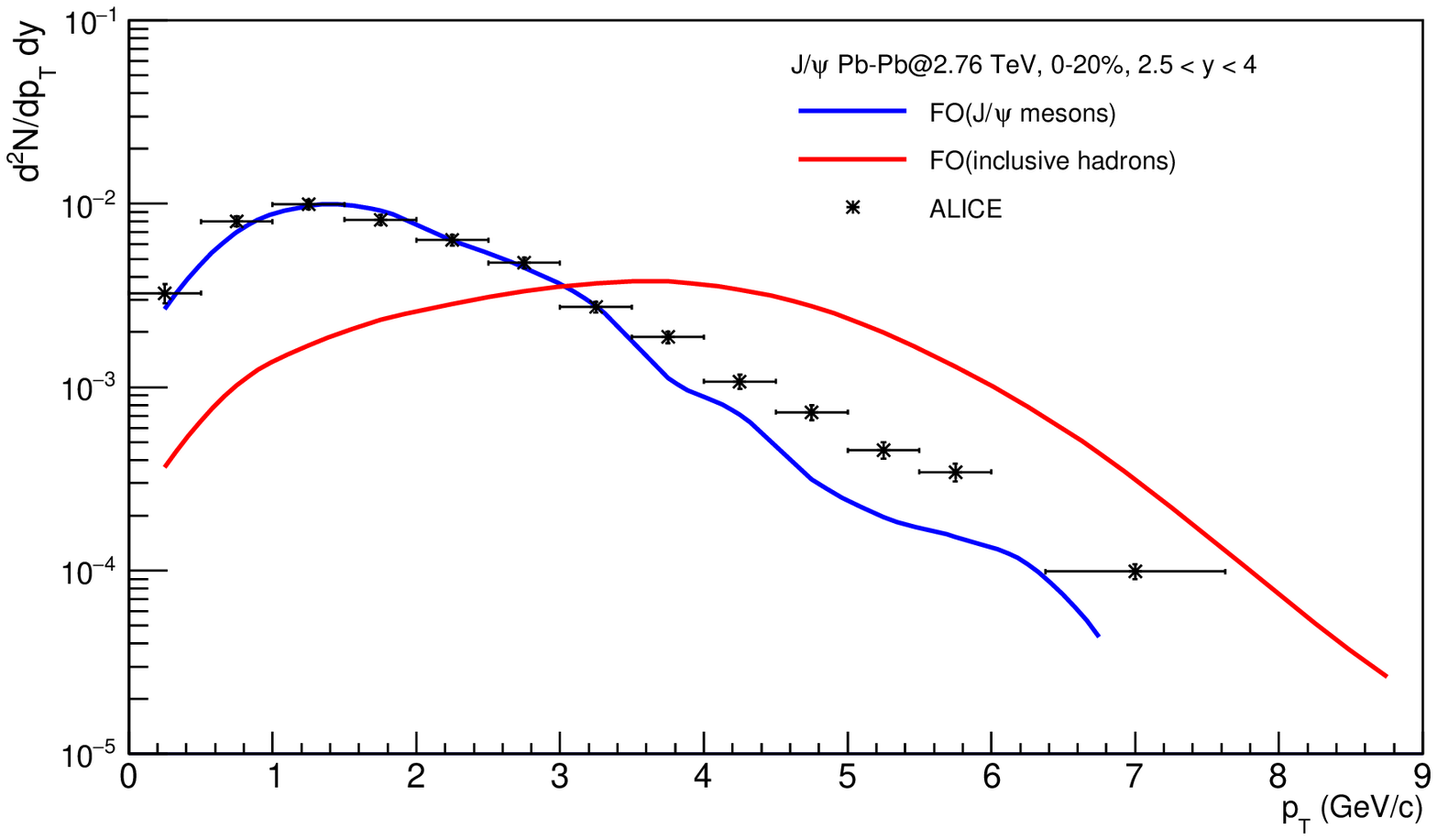}
\caption{\label{jpsi_temp}
Transverse momentum spectrum of $J/\psi$-mesons at rapidity $2.5<y<4$ in 
20\% of most central PbPb collisions at $\sqrt{s_{\rm NN}}=2.76$ TeV. The points are ALICE 
data~\cite{Adam:2015isa}, histograms are simulated HYDJET++ events (red -- freeze-out
parameters as for inclusive hadrons, blue -- ``early'' thermal freeze-out).}
\end{minipage}
\hspace{2pc}
\begin{minipage}{18pc}
\includegraphics[width=18pc]{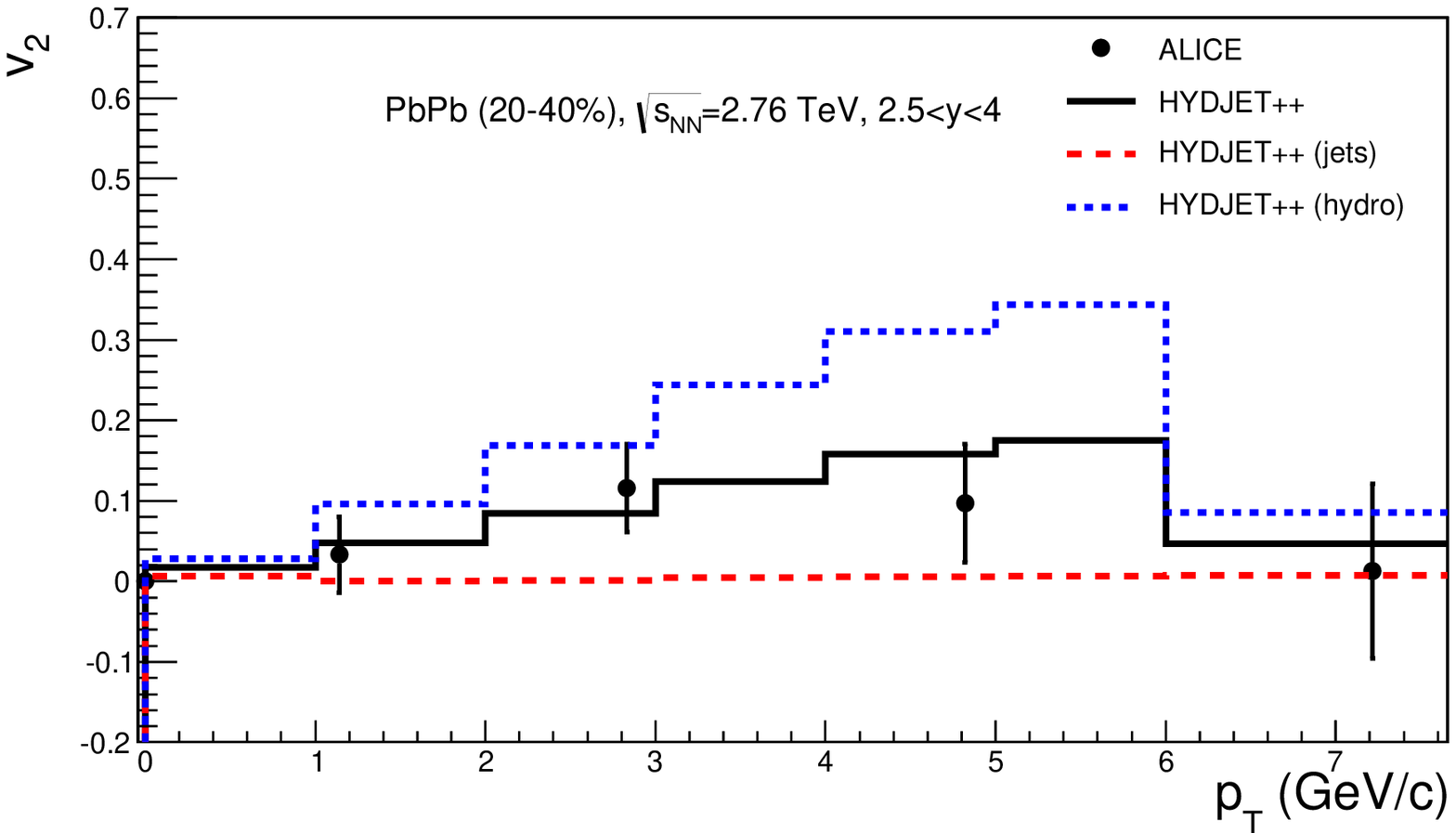}
\caption{\label{jpsi_v2} Elliptic flow coefficient $v_2(p_{\rm T})$ of $J/\psi$-mesons at rapidity
$2.5<y<4$ for 20--40\% centrality of PbPb collisions at $\sqrt s_{\rm NN}=2.76$ TeV. 
The points are ALICE data~\cite{ALICE:2013xna}, histograms are simulated HYDJET++ events 
(blue -- soft component, red -- hard component, black - both components).}
\end{minipage} 
\end{figure}

Some time ago it was demonstrated in~\cite{Lokhtin:2010ze} that if $J/\psi$-mesons are produced in HYDJET++ at 
the same freeze-out parameters as for inclusive (light) hadrons, then simulated $p_{\rm T}$- and 
$y$-spectra in 20\% of most central AuAu collisions at RHIC energy $\sqrt{s_{\rm NN}}=200$ GeV are much wider than ones measured by PHENIX~\cite{Adare:2006ns}. However the assumption that the thermal 
freeze-out for $J/\psi$-mesons happens at the same temperature as chemical freeze-out (with reduced 
collective velocities) allows HYDJET++ to reproduce the data. It turned out that the similar situation holds at the LHC. Figure~\ref{jpsi_temp} shows the comparison of HYDJET++ simulations 
with the ALICE data~\cite{Adam:2015isa} for $p_{\rm T}$-spectrum of $J/\psi$-mesons in 20\% of most 
central PbPb collisions at $\sqrt{s_{\rm NN}}=2.76$ TeV. One can see that if thermal freeze-out for 
$J/\psi$'s happens at the same temperature as chemical freeze-out (with reduced collective 
velocities), then simulated spectra match the data up to $p_{\rm T} \sim 3$ GeV/$c$. The discrepancy 
at high $p_{\rm T}$ may indicate on the necessity to tune PYTHIA for
charmonium production in this kinematic domain. The $p_{\rm T}$-dependence of the elliptic flow coefficient $v_2$ measured by ALICE~\cite{ALICE:2013xna} is also reproduced by HYDJET++
simulations in such a case (figure~\ref{jpsi_v2}). The contribution of non-thermal component is
important at high $p_{\rm T}$.

Thus we conclude that $J/\psi$-meson production in heavy ion collisions at RHIC and LHC may be
reproduced by HYDJET++ model under the assumption that thermal freeze-out of charmonia happens 
appreciably before thermal freeze-out of light hadrons~\cite{Bugaev:2001sj}, presumably at chemical
freeze-out (with reduced radial and longitudinal collective velocities). Our interpretation of this 
is that the significant part of $J/\psi$-mesons (up to $p_{\rm T} \sim 3$ GeV) is thermalized, but 
without being in a kinetic equilibrium with the medium.

\begin{figure}[h]
\begin{minipage}{18pc}
\includegraphics[width=18pc]{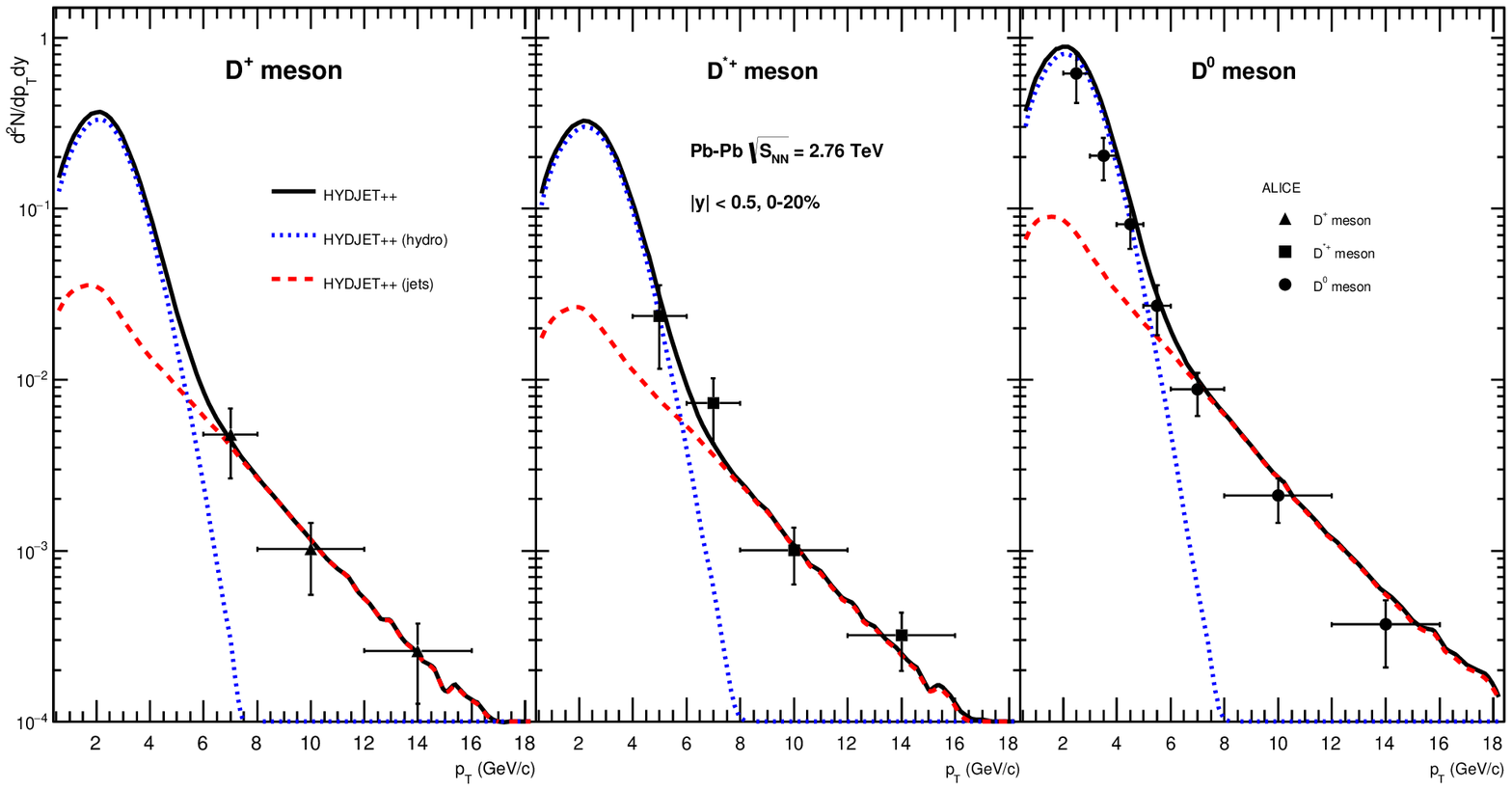}
\caption{\label{d_pt}
Transverse momentum spectra of D-mesons at rapidity $\mid y \mid < 0.5$
in 20\% of most central PbPb collisions at $\sqrt{s_{\rm NN}}=2.76$ TeV. The points are ALICE 
data~\cite{ALICE:2012ab}, histograms are simulated HYDJET++ events (blue -- soft component, red -- hard component, black - both components).}
\end{minipage}
\hspace{2pc}
\begin{minipage}{18pc}
\includegraphics[width=18pc]{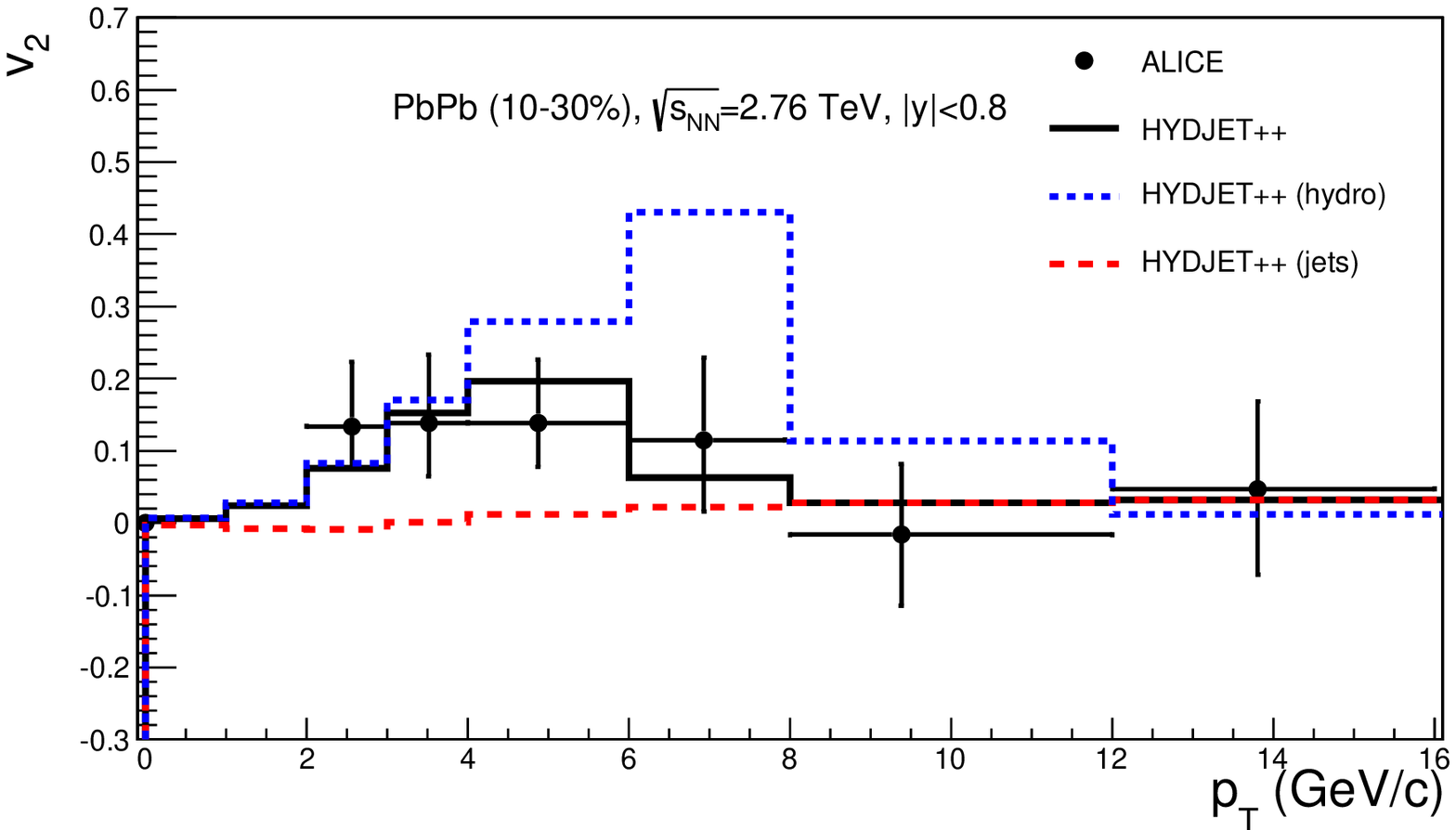}
\caption{\label{d_v2} 
Elliptic flow coefficient $v_2(p_{\rm T})$ of D$^0$-mesons at rapidity
$\mid y \mid <0.8$ for 10--30\% centrality of PbPb collisions at 
$\sqrt s_{\rm NN}=2.76$ TeV. The points are ALICE data~\cite{Abelev:2014ipa}, histograms are 
simulated HYDJET++ events (blue -- soft component, red -- hard component, black - both components).}
\end{minipage} 
\end{figure}

\section{D-meson production in PbPb collisions at the LHC}

At first we have checked that $p_{\rm T}$-spectrum of D-mesons measured by 
STAR~\cite{Adamczyk:2014uip} in 10\% of most central AuAu collisions at $\sqrt{s_{\rm NN}}=200$ GeV
is reproduced by HYDJET++ simulation with fugacity $\gamma_c=7$ and the same freeze-out parameters as for $J/\psi$-mesons (but not for inclusive hadrons). It means that D-mesons like $J/\psi$-mesons are not in a kinetic equilibrium with the medium at RHIC. However the situation gets changed dramatically at the LHC. Figure~\ref{d_pt} shows the comparison of HYDJET++ simulations   
with the ALICE data~\cite{ALICE:2012ab} for $p_{\rm T}$-spectra of D$^0$, D$^{\pm}$ and D$^{*\pm}$ 
mesons in 20\% of most central PbPb collisions at $\sqrt{s_{\rm NN}}=2.76$ TeV. The freeze-out 
parameters as for inclusive hadrons have been used, the fugacity value $\gamma_c=11.5$ being fixed 
from $J/\psi$ yield. The model results are close to the data within the experimental  
uncertainties. Note that $p_{\rm T}$-spectra and nuclear modification factors $R_{\rm AA} 
(p_{\rm T})$ (the ratio of particle yields in AA and pp collisions normalized on the mean number of 
NN sub-collisions) of D-mesons are reproduced by HYDJET++ up to very high $p_{\rm T}$. 
So the treatment of heavy quark energy loss in hard component of the model (PYQUEN) seems quite
successful. In addition, HYDJET++ is able to reproduce the 
ALICE data ~\cite{Abelev:2014ipa} on $p_{\rm T}$-dependence of the elliptic flow coefficient $v_2$ (figure~\ref{d_v2}). 
 
Thus in contrast to RHIC, thermal freeze-out of D-mesons at the LHC happens
simultaneously with thermal freeze-out of light hadrons. Therefore the significant part
of D-mesons (up to $p_{\rm T} \sim 4$ GeV/$c$) seems to be in a kinetic equilibrium with the medium.
 
\section{Summary}

Momentum spectra and elliptic flow of D and $J/\psi$ mesons in PbPb collisions at the LHC are reproduced by two-component model HYDJET++ including thermal (soft) and non-thermal (hard) 
components. Thermal freeze-out of D-mesons happens simultaneously with thermal freeze-out of light hadrons. Thermal freeze-out of $J/\psi$-mesons happens appreciably before freeze-out of light hadrons, presumably at chemical freeze-out (with reduced radial and longitudinal collective velocities). 
Thus the significant part of D-mesons (up to $p_{\rm T} \sim 4$ GeV/c) seems to be in a kinetic equilibrium with the medium, while $J/\psi$ mesons are not. The production of charmed mesons at high transverse momenta is determined by non-thermal charm production mechanism and medium-induced heavy quark energy loss. 

\section{Acknowledgments}

Discussions with L.V. Bravina, V.L. Korotkikh, L.V. Malinina, A.M. Snigirev, E.E. Zabrodin and 
S.V. Petrushanko are gratefully acknowledged. We thank our colleagues from ALICE and CMS 
collaborations for fruitful cooperation. This work was supported by the Russian Scientific Fund 
under Grant No. 14-12-00110 in a part of computer simulation of $p_{\rm T}$-spectrum and elliptic flow of charmed mesons in PbPb collisions. G.E. acknowledges the European Social Fund within the framework of realizing the project Support of Inter-sectoral Mobility and Quality
Enhancement of Research Teams at Czech Technical University in Prague, CZ.1.07/2.3.00/30.0034.

\end{document}